\pdfoutput=1
\documentclass{phep}       

\journal{JAIS-ID}
\vol{xx}

\received{xx January 2025}
\published{xx March 2025}

\usepackage{amsmath}
\usepackage{amssymb}
\usepackage[hidelinks]{hyperref}

\def\be{\begin{equation}}
\def\ee{\end{equation}}
\def\bea{\begin{eqnarray}}
\def\eea{\end{eqnarray}}

\renewcommand{\thefootnote}{\fnsymbol{footnote}}

\begin{document}

\title{Portable Glass-RPC Telescope for Muography and
Future Applications at Laser Wake Field Accelerators}

\author{
S. Ikram,\auno{1}\footnotemark\ 
D. Ahmadi,\auno{2}
U. Chaulagain,\auno{3}
A. Cimmino,\auno{3}
E. Cortina Gil,\auno{1} 
P. Demin,\auno{1} 
A. Giammanco,\auno{1} 
G. M. Grittani,\auno{3}
M. Lagrange,\auno{1}
M. Nevrkla,\auno{3,4}
J. Sisma,\auno{3,4}
M. Tytgat,\auno{2}
R. Versaci,\auno{3} 
and F. Vitha\auno{3,4}
}

\address{
\begin{flushleft}
    
$^1$Centre for Cosmology, Particle Physics and Phenomenology (CP3), Universit\'e catholique de Louvain, Louvain-la-Neuve, Belgium\\
$^2$Inter-University Institute for High Energies (IIHE), Vrije Universiteit Brussel, Brussels,
Belgium\\
$^3$ELI Beamlines Facility, The Extreme Light Infrastructure ERIC, Dolni Brezany,
Czechia\\
$^4$Faculty of Nuclear Sciences and Physical Engineering, Czech Technical University in Prague,  Prague, Czechia\\
\end{flushleft}
}

\begin{abstract}

We report a 2025 beam test at ELBA, the multi-GeV laser wakefield accelerator at ELI Beamlines, using a Resistive Plate Chamber (RPC) telescope with an active area of $16 \times 16~{cm}^2$. The portable design supports autonomous operation in challenging environments. Previous cosmic-ray measurements established its tracking capability for muography. During the campaign at ELBA, $30~{fs}$ long, $0.5~{PW}$  laser pulses interacted with a plasma target at a $0.2~{Hz}$  repetition rate, where a tungsten converter upstream of the beam dump enhanced muon production. This was the first deployment of an RPC telescope in a high-power laser environment. The detector remained operational in the presence of radiation and electromagnetic interference. Although parasitic running and limited statistics prevented definitive identification of beam-induced muons, the measurements offered insight into background activity and showed evidence of reconstructed tracks. They establish the feasibility of a dedicated campaign with controlled beam conditions and improved muon sensitivity. This contribution forms part of the International Muography Workshop 2026 proceedings.

\par\vspace{8pt}


\par\vspace{4pt}


\end{abstract}

\maketitle


\renewcommand{\thefootnote}{\fnsymbol{footnote}}
\footnotetext{Corresponding author: sumaira.ikram@uclouvain.be}

\section{Introduction}
Muography uses naturally occurring cosmic-ray muons to image large or dense objects through absorption or scattering without damaging it~\cite{Bonechi_2020}. Muons penetrate matter and carry information about the material they traverse. Measurements typically rely on either the reduction in muon flux through the target or changes in muon trajectories caused by multiple Coulomb scattering. These approaches serve applications ranging from maritime security inspections and volcanic monitoring to the investigation and preservation of cultural heritage objects, as well as the potential detection and monitoring of CO$_2$ plumes in geological formations~\cite{Basiri2026CO2}. The quality of the reconstructed image depends on the detector arrangement and its ability to detect muons and measure their positions accurately. Practical deployment imposes further constraints. Systems intended for remote or difficult-to-access sites must combine adequate tracking performance with portability, autonomous operation, and resistance to demanding environmental conditions~\cite{Vishal_kumar2024}. Our project develops compact, robust, and autonomous muon telescopes for measurements in confined, remote, or logistically challenging environments~\cite{Basnet_2020}. The system employs sealed glass Resistive Plate Chambers (RPCs), chosen for their tracking capabilities, cost effectiveness, and ease of construction~\cite{ MuographyBook}. Sealed operation limits dependence on external gas supplies and improves safety during field deployment~\cite{gamage2022}.
Cosmic-ray muons require no artificial source, but their low and uncontrolled flux often leads to long measurement times. Laser-plasma accelerators offer a potential alternative by producing short, directional bursts of multi-GeV electrons~\cite{Rockafellow2025}, which generate secondary particles, including muons, when directed onto high-density materials~\cite{Zhang2025}. Such sources offer controlled pulse timing and higher directional flux, potentially reducing acquisition times, particularly for imaging small objects of interest. This approach remains at an early experimental stage, with muon yield, background suppression, and detector operation near high-power lasers still under investigation~\cite{Terzani2025}.
In 2025, our team deployed the telescope at the ELBA laser wakefield electron accelerator at ELI Beamlines to study highly penetrating charged particles generated by  $30~{fs}$  long, $0.5~{PW}$ laser pulses operating at $0.2~{Hz}$ repetition rate~\cite{Sisma2026}. A tungsten target was placed upstream of the concrete beam dump to enhance muon production. This was the first deployment of a gas-tight RPC telescope in a high-power laser environment. Despite limited statistics and parasitic data-taking conditions, the detector operated stably and provided initial measurements of the background and tracking performance. A detailed description of the experimental campaign is presented in Ref.~\cite{LPA-Ikram2026}. While this proceedings contribution focuses and expands on operational feasibility and background characterization, Reference~\cite{LPA-Ikram2026} also provides FLUKA Monte Carlo fluence estimates, formal upper-limit statistical calculations, and controlled laboratory validation tests.

Section~\ref{sec:rpc-design} describes the RPC design and laboratory setup. Section~\ref{sec:eli_delpoyment} presents the detector deployment at ELI and an overview of the data-taking campaign and  different shielding configurations. Section~\ref{sec:Event_selection} presents the event-selection criteria and the algorithm used to identify and reconstruct particle tracks. Section~\ref{sec:results} presents the main results and their discussion. Finally, Section~\ref{sec:summary} summarizes the main findings and outlines future work.


\section{Glass-RPC Design and Laboratory setup}
\label{sec:rpc-design}

\begin{figure}
    \centering
    \includegraphics[width=0.45\textwidth]{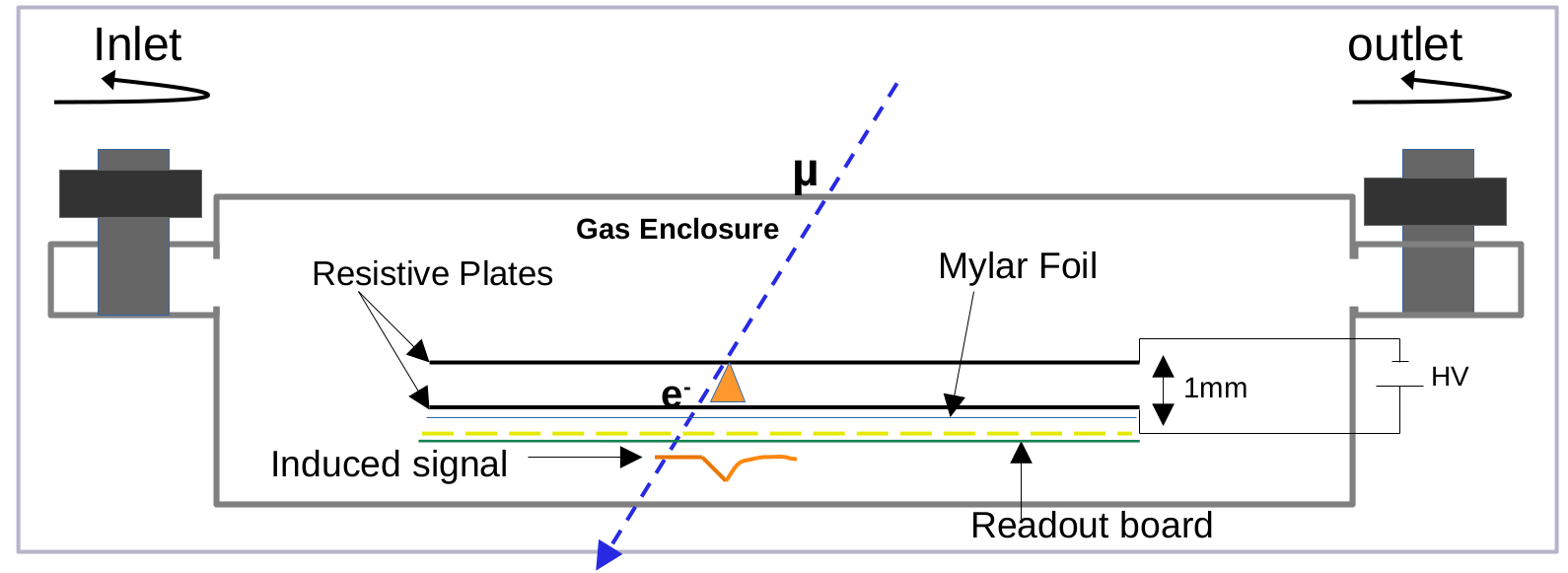}
    \caption{Illustration of the glass RPC detector integrated within a sealed gas-tight housing.
}
    \label{schematic}
\end{figure}

\begin{figure}
    \centering
    \includegraphics[width=0.4\textwidth]{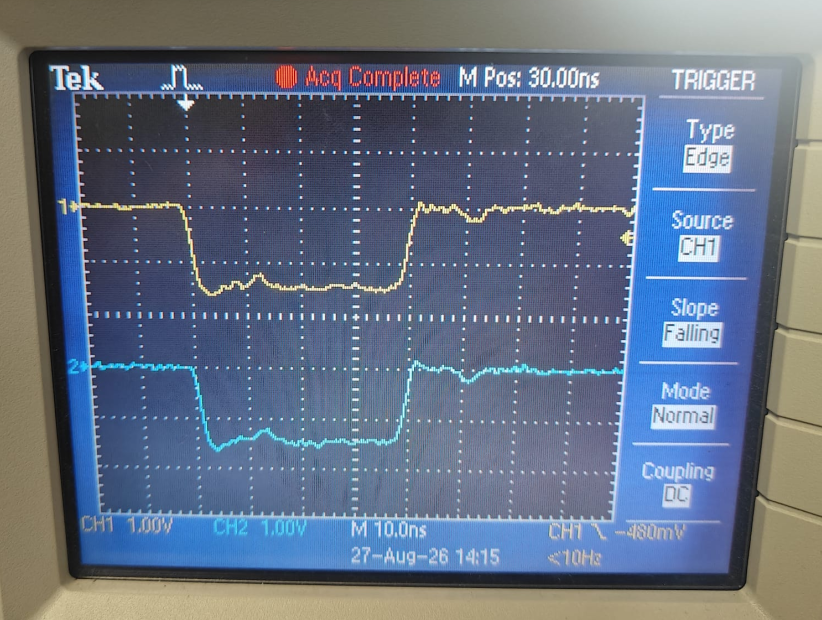}
    \caption{Oscillogram of the two scintillator logic signals used to form the coincidence trigger. Both pulses have an approximate width of $40~ns$. The horizontal scale is $10~ns$ per division.
}
    \label{coincidence_window}
\end{figure}

The detector setup comprises of three RPCs, consist of $20 \times 20~{cm}^2$ glass plates with a thickness of $1.1~mm$. The central $16 \times 16 ~{cm}^2$ region of the plates defines the active area and is coated with a resistive paint having a surface resistivity of 0.5 to 1.0 $M\Omega/\square$. A $1~mm$ gas gap is maintained uniformly between the glass plates. 
Each RPC together with its readout board is housed inside an aluminum gas enclosure filled with a gas mixture of 95.2\% Freon, 0.3\% SF$_6$, and 4.5\% isobutane at atmospheric pressure. The schematic of the chambers is shown in Figure~\ref{schematic}. The enclosure has an internal volume of 1.9 $L$ and dimensions of about $32.5\times24.0\times2.45~{cm}^3$. Before operation, each chamber is flushed for four to five volume exchanges at a flow rate of  0.2 $L/min$ to properly remove residual gas. The inlet and outlet are then closed, allowing the chambers to work in sealed mode. The present readout board contains 16 strips, each $18~cm$ long and $0.9~cm$ wide, with a strip pitch of $1~cm$~\cite{RPC-Ikram2025}. The trigger is generated by the logical AND of the two scintillator signals after pulse shaping by the quad timer NIM module. A coincidence occurs when the two logic pulses overlap, with their widths defining the coincidence acceptance window. The oscillogram show negative-going pulses approximately $40~ns$ wide at a horizontal scale of $10~ns$ per division as in Figure~\ref{coincidence_window}. The trigger time corresponds to the leading edge of the coincidence output, which is a falling voltage edge for negative NIM logic.

The detector data-acquisition (DAQ) is based on charge-sensitive front-end electronics originally developed for the CMS Muon-RPC system~\cite{Abbrescia:2000sq}. The board combines a charge-sensitive ASIC with an FPGA for signal acquisition and digital processing. The RPC operating voltage is supplied by a dedicated HV system. Further details of front-end electronics and DAQ architecture are reported in Ref.~\cite{Donya_2026}. Detector performance was studied for several combinations of applied high voltage and charge threshold.

\section {ELI Deployment and Data Taking}
\label{sec:eli_delpoyment}
In 2025, we performed three beam-test campaigns in April/May, August, and September. During these campaigns, we conducted a series of beam and cosmic runs using different shielding configurations and detector positions relative to the beam source. During these data-taking periods, we had no control over the beam parameters, as other groups were deciding on the beam parameters and we were using the beam parasitically. Measurements were grouped into three categories of data-taking: D1, D2, and D3.
\begin{itemize}
\item \textbf{D1:} Data were collected over two weeks during April and May in the ELBA experimental hall as shown in Figure~\ref{fig:detectorsetup_may}. The detector was positioned at beam height, 335~cm from the beam dump, and equipped with polyethylene and lead shielding. Layer~2 was oriented perpendicular to layers~1 and~3 in vertical orientation. The position difference of hits between layers~1 and~3 was used to distinguish cosmic muons from beam-induced muon candidates, while layer~2 helped to provide information on the position of the beam-induced particle-cone. Two datasets were recorded: D1 \textit{cosmic}, with the beam off, and D1 \textit{beam}, with the beam on under different shielding conditions.

\item \textbf{D2:} During the August campaign, the detector was positioned $18~m$ downstream of the beam dump, as shown in Figure~\ref{fig:detectorsetup_august}. All three RPC chambers were aligned in the same orientation vertically, allowing track reconstruction in XZ plane. Three datasets were collected: D2 \textit{cosmic}, with the beam off and no shielding; D2 \textit{beam}, with the beam on, both without shielding and with $5~cm$ of lead followed by $14~cm$ of polyethylene; and D2 \textit{tag}, with the beam on and the laser trigger signal recorded. Additional polyethylene shielding was placed around the detector to reduce neutron and back-scattered backgrounds.

\item \textbf{D3:} Data were collected during August and September at the GAMMATRON beamline~\cite{Chaulagain2022,Chaulagain2025}. The electron beam energy reached up to $1~GeV$, where muon production is expected to be negligible. This dataset therefore served as a control sample for beam-induced backgrounds. The detector was positioned approximately $9~m$ behind a thick concrete wall, and several lead and polyethylene shielding configurations were studied, as described in Section~\ref{sec:shielding}. More details can be found in Ref.~\cite{LPA-Ikram2026}.
\begin{figure}[h!]
    \centering

    \begin{tabular}{cc}
        \includegraphics[width=0.47\linewidth]{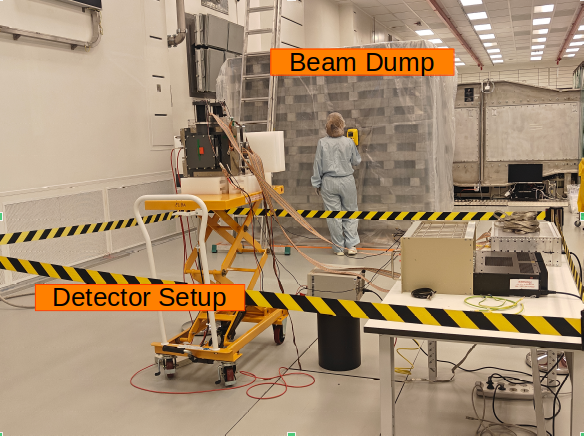} &
        \includegraphics[width=0.41\linewidth]{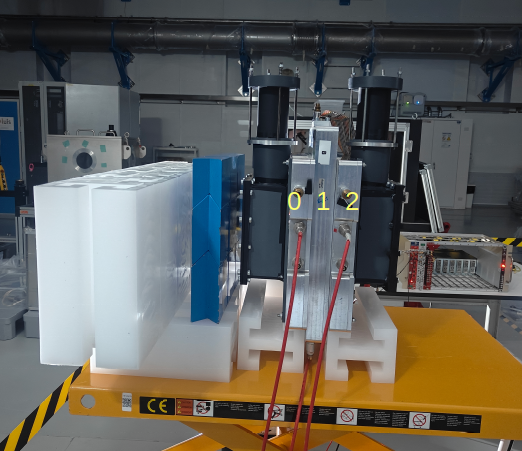} \\
        (a) & (b)
    \end{tabular}

    \caption{(a) RPC telescope installed in the experimental hall during the April/May campaign at a distance of $335~cm$ from the beam dump; (b) detector arrangement surrounded by lead and polyethylene shielding~\cite{LPA-Ikram2026}.}
    \label{fig:detectorsetup_may}
\end{figure}

\begin{figure}
    \centering
    \begin{tabular}{cc}
    \includegraphics[width=0.54\linewidth]{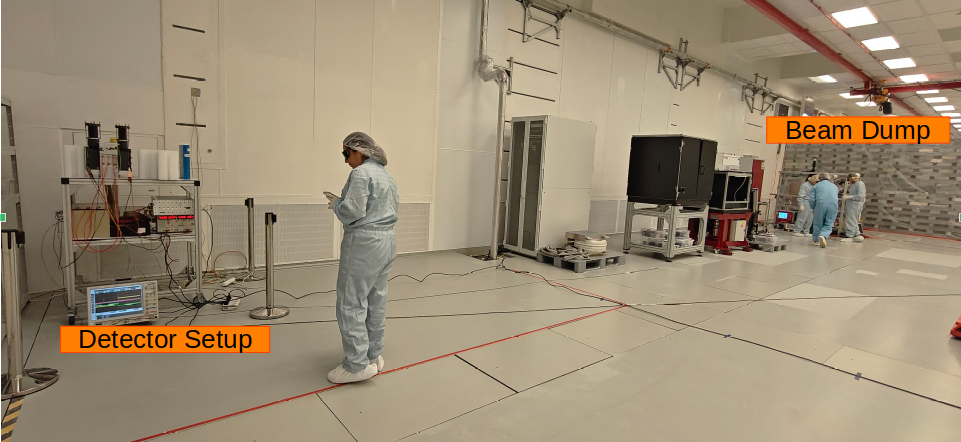} &
    \includegraphics[width=0.4\linewidth]{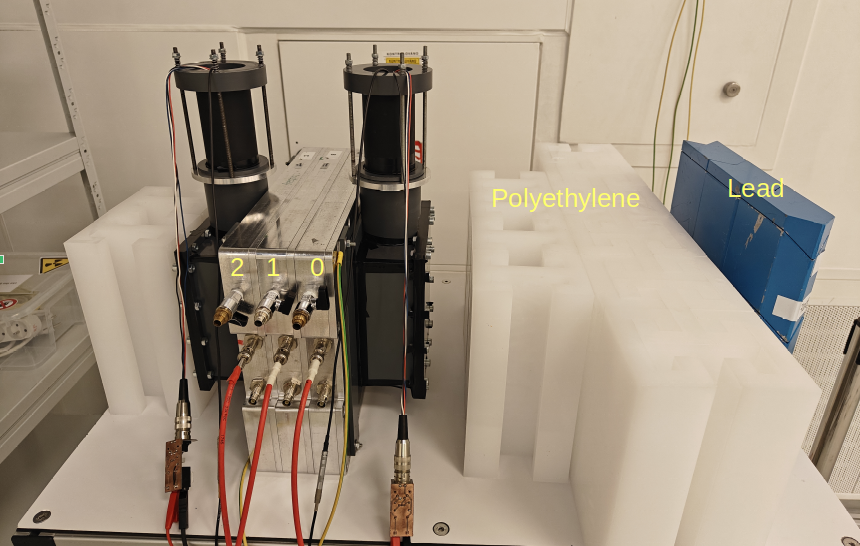} \\
    (a) & (b)
    \end{tabular}
    \caption{(a) RPC telescope installed in the experimental hall during the August campaign at a distance of $18~m$ from the beam dump; (b) all detectors arranged in the same direction, surrounded by lead and polyethylene shielding~\cite{LPA-Ikram2026}.}
    \label{fig:detectorsetup_august}
\end{figure}

\end{itemize}

\section{Event selection and track identification}
\label{sec:Event_selection}
The cosmic runs are taken as the baseline. During cosmic events, mostly only one strip was fired per event in each chamber. In contrast, beam events had higher multiplicity and occupancy. An event display for each data is shown in  Figure~\ref{fig:event_displays}. 
A simple algorithm is used to identify muon tracks and determine their direction relative to the horizontal axis. Data acquisition uses a $200~MHz$ FPGA clock, corresponding to 250 samples with time bin width of $5~ns$. Spatially adjacent fired strips within the same $5~ns$ time bin are grouped into clusters, with the cluster position defined by the strip centroid. Events are categorized as Diamond, Platinum, Gold, Silver, Bronze, or Other based on hit multiplicity and timing.

\begin{figure}
    \centering
    \includegraphics[width=0.9\linewidth]{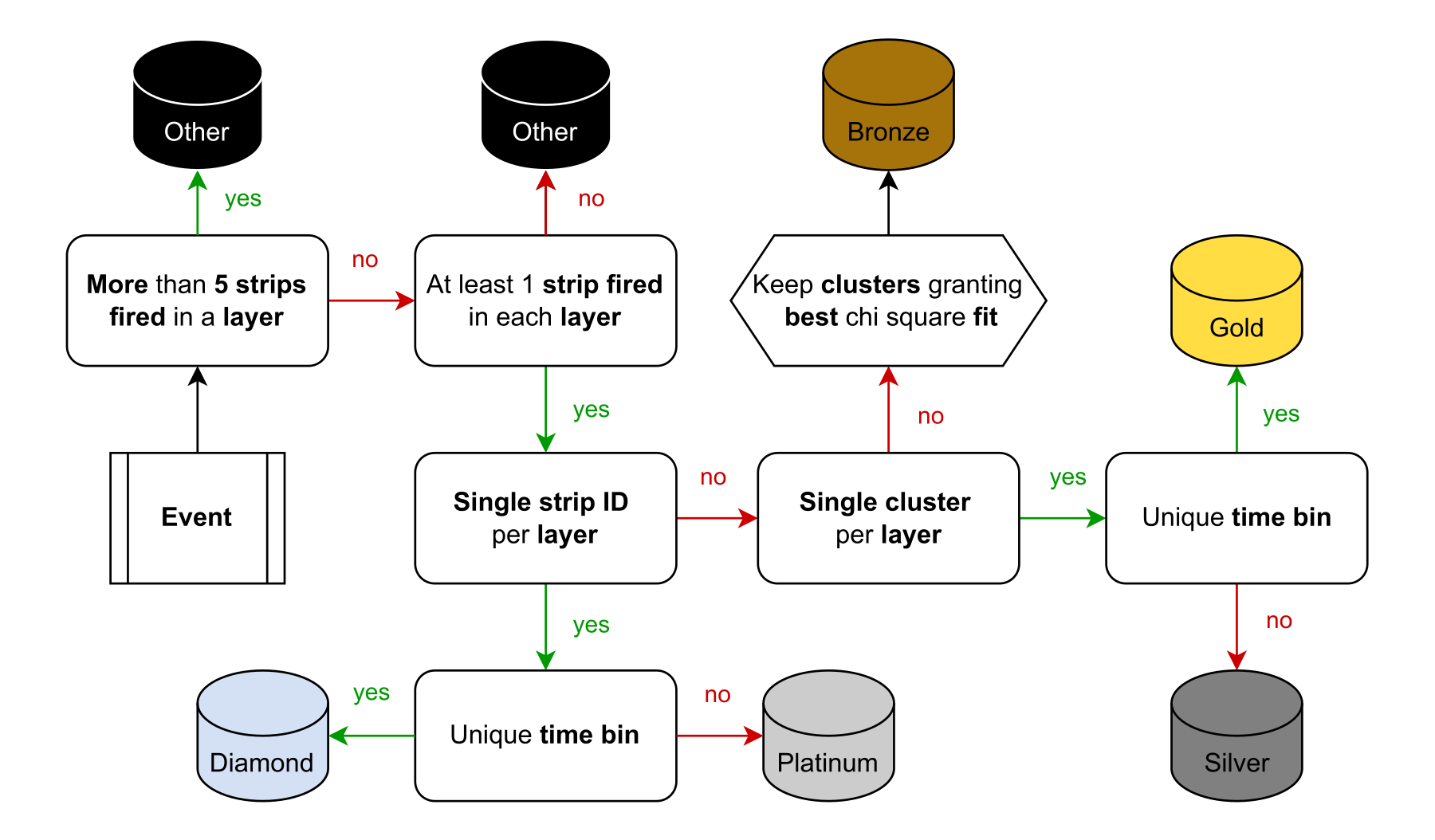}
    \caption{Decision tree used for track reconstruction in the D2 and D3 datasets. For D1, Bronze events are reconstructed without applying the $\chi^2$ based selection~\cite{LPA-Ikram2026}.}
    \label{fig:decision_tree}
\end{figure}

\begin{itemize}
\item \textbf{Diamond / Platinum}: Exactly one strip per layer is fired, labeled as diamond in the same time bin or platinum if the fired strips are in different time bins.
\item \textbf{Gold/ Silver}: Exactly one cluster per layer, gold in the same and silver if in different time bins.
\item \textbf{Bronze}: Multiple clusters are present; for D2 and D3, the best track is selected using a  $\chi^2$-based criterion.
\item \textbf{Others}: Events with more than 5 hits and with hits in only two RPC chambers are excluded from tracking.
\end{itemize}
 The classification scheme is shown in  Figure \ref{fig:decision_tree}.

\section{Results and Discussion}
\label{sec:results}
\begin{figure}
    \centering
    \includegraphics[width=0.4\textwidth]{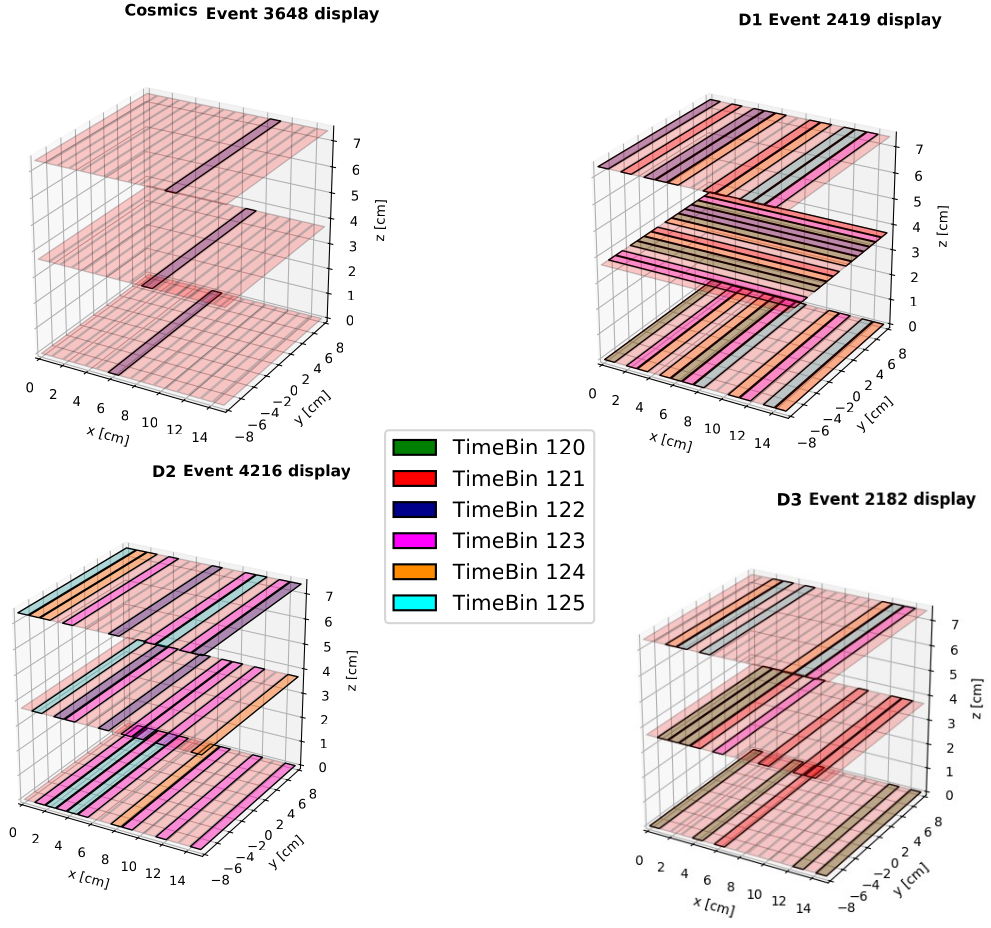}
   \caption{Representative event displays for the cosmic-ray, D1, D2, and D3 data samples. Top left: cosmic-ray event. Top right: D1 event. Bottom left: D2 event. Bottom right: D3 event.}
    \label{fig:event_displays}
\end{figure}

This section presents the data recorded during the D1, D2, and D3 campaigns. A chamber operating voltage of $7~kV$ was selected through laboratory cosmic-ray measurements to maximize detection efficiency while keeping the leakage current below $60~nA$ at front-end electronic thresholds of 30, 35, and 40. During the D1 beam test, layer 2 was oriented orthogonally so its signal distribution would help verify the detectors’ horizontal alignment with the core, rather than the edge, of the particle cone produced downstream of the beam dump.
The analysis aims to assess the detector’s sensitivity to beam-produced muons. Strip occupancy, strip multiplicity, and angular distributions are compared across the cosmic and beam datasets, followed by an examination of track properties. The timing structure of the raw detector data provides further insight into the nature of the recorded signals. 
Figure~\ref{fig:event_displays} shows representative event displays for the four data samples: cosmic rays, D1, D2, and D3. In cosmic-ray events, typically only one strip is fired in each detector layer. The D1 and D2 samples were recorded with higher-energy beams and exhibit larger strip multiplicities. In contrast, D3 was recorded with a beam energy below $1~GeV$ and shows fewer fired strips than D1.

\subsection {Occupancy and Multiplicity}

\begin{figure}[htbp]
    \centering

    \includegraphics[
        width=\linewidth,
        height=0.36\textheight,
        keepaspectratio
    ]{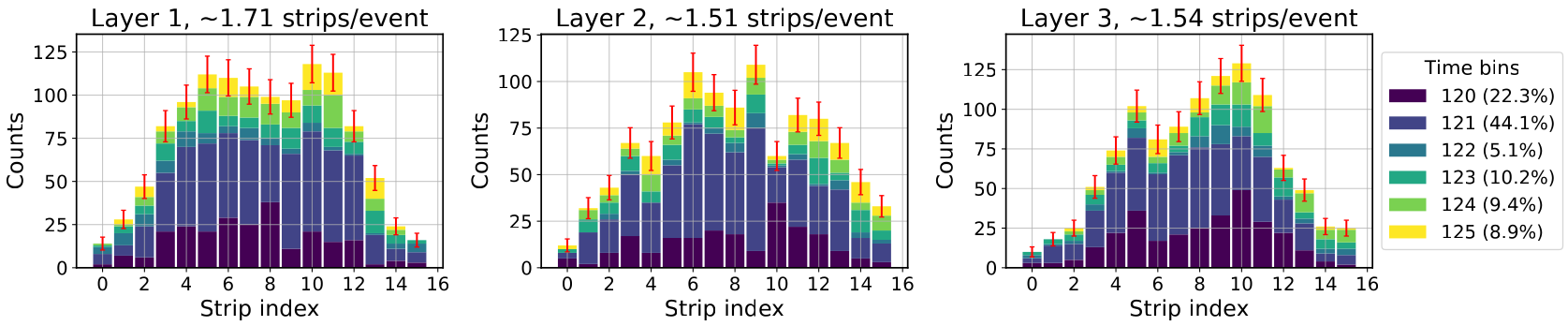}

    \vspace{0.2cm}

    \includegraphics[
        width=\linewidth,
        height=0.36\textheight,
        keepaspectratio
    ]{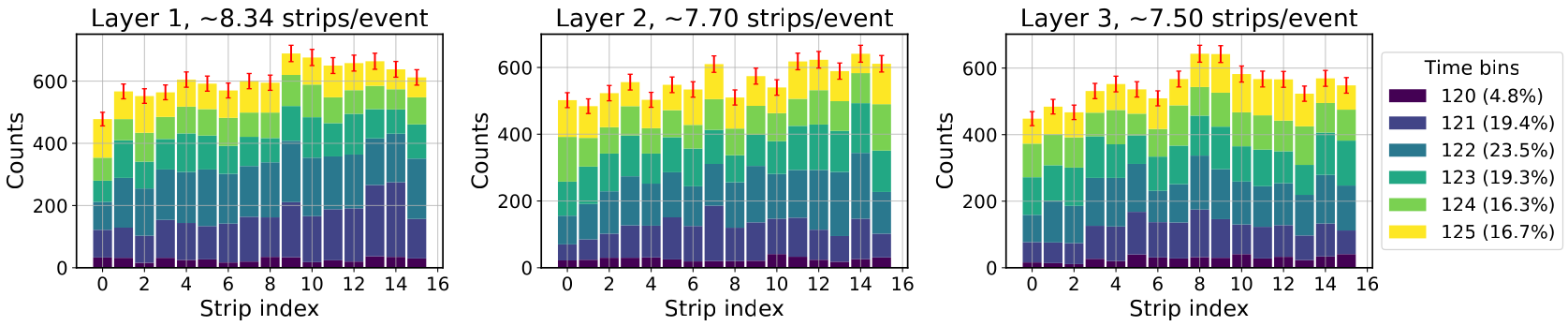}

    \caption{Strip occupancy distributions for the D1 dataset.
    The upper panel shows cosmic-ray data, while the lower panel
    shows beam data with polyethylene followed by lead (PE--Pb).
    Poisson uncertainties are described in Ref.~\cite{LPA-Ikram2026}.}
    \label{fig:D1_occupancy}
\end{figure}

\begin{figure}[h!]
    \centering

    \begin{tabular}{cc}
        \includegraphics[width=0.45\linewidth]{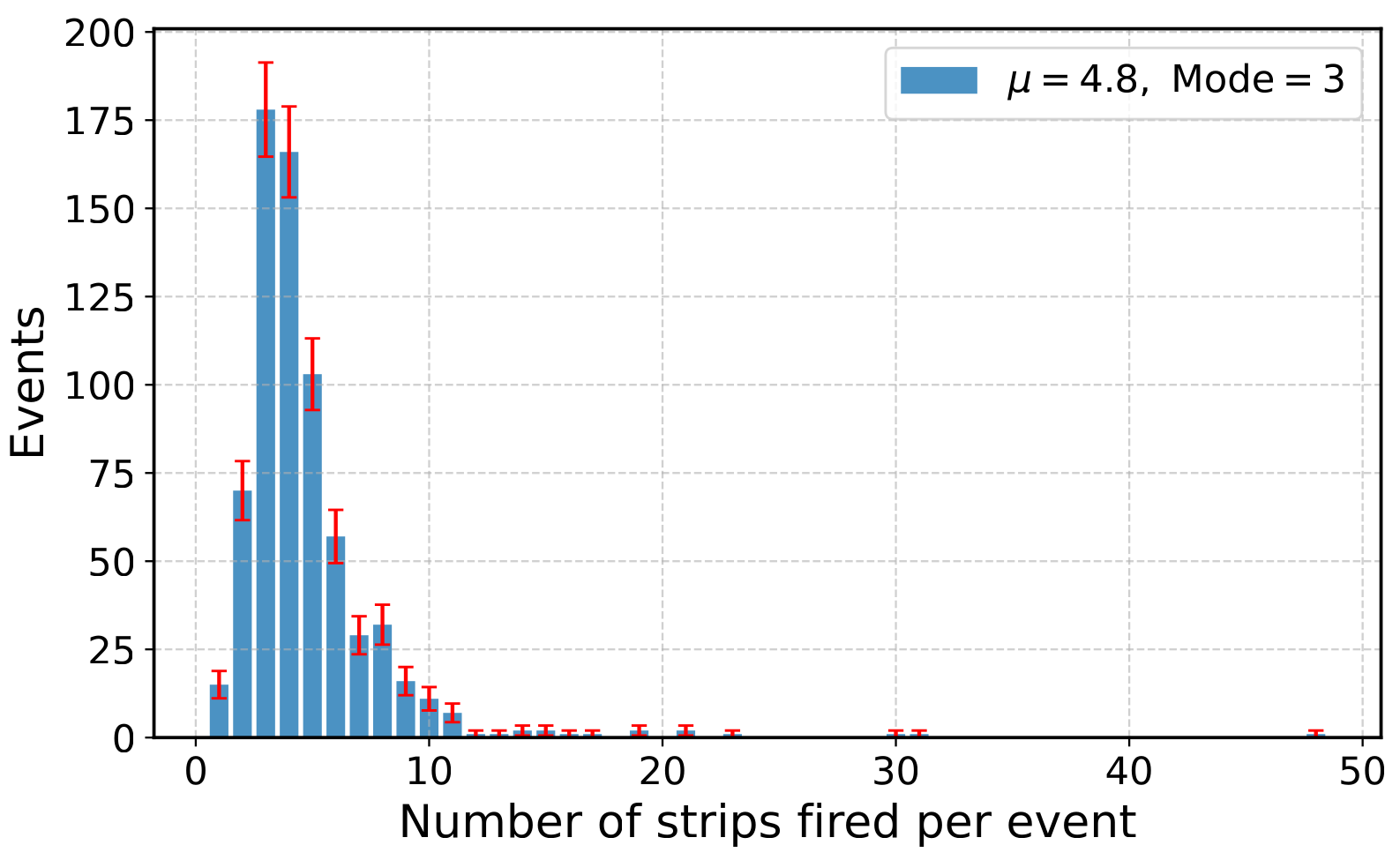} &
        \includegraphics[width=0.45\linewidth]{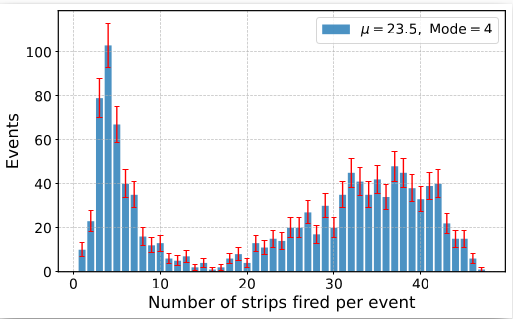} 
    \end{tabular}

    \caption{Distributions of the number of fired strips per event for the (left) D1 cosmics and (right)D1 beam data. The legend lists the mode and mean $\mu$  of each distribution. Poisson uncertainties are indicated in red~\cite{LPA-Ikram2026}.}
    \label{D1_multiplicity}
\end{figure}

\begin{figure}
    \centering
    \includegraphics[width=0.48\textwidth]{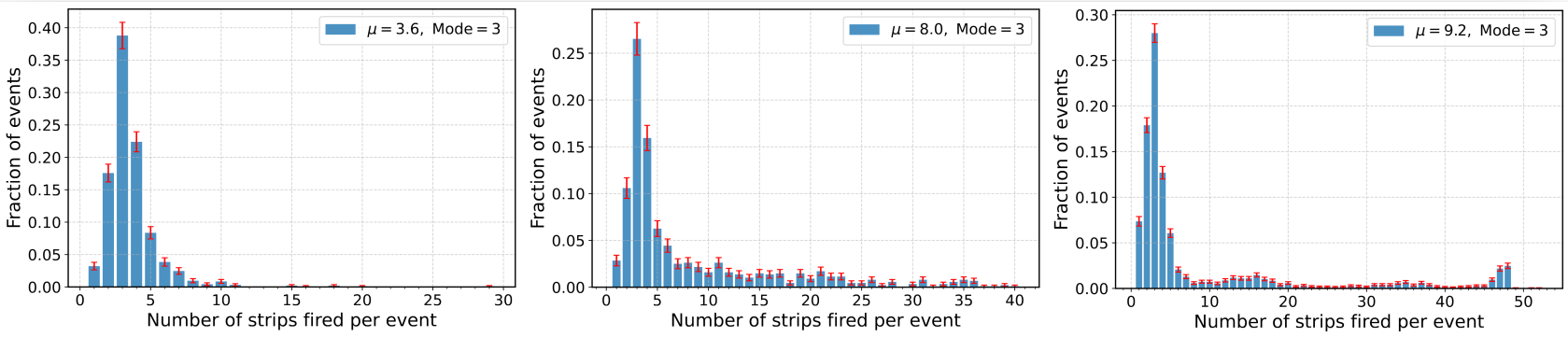}
    \caption{Distributions of the number of fired strips per event for the D2 cosmic (left), D2 beam (center), and D3 beam (right) datasets. The legend lists the mode and mean $\mu$  of each distribution. Poisson uncertainties are indicated in red~\cite{LPA-Ikram2026}.
}
    \label{D2_D3_mutiplicity}
\end{figure}

Figure~\ref{fig:D1_occupancy} compares the strip occupancy and timing distributions of the D1 cosmic reference run and beam datasets. The analysis selects bins 120–125 ($605–630~ns$), corresponding to the RPC signal window relative to the trigger. The cosmic run provides a baseline for the detector response to single muons, with most hits concentrated in the $5~ns$-wide bin 121, corresponding to $605~ns$ after the scintillator trigger. Beam activity spans bins 121–123 ($605–615~ns$), with higher strip multiplicities and broader timing distributions.
Figure~\ref{D1_multiplicity} shows a mean strip multiplicity of 23 for D1 beam data, compared with 4.8 for cosmic data. A Mode of 3 fired strips is consistent with cosmic-like events, while the beam distributions extend to approximately 50 fired strips. The higher multiplicities and occupancy differences across absorber configurations are consistent with additional beam-induced secondary activity, although strip multiplicity alone does not determine particle numbers or origins.
\begin{figure}
    \centering
    \includegraphics[width=0.45\textwidth]{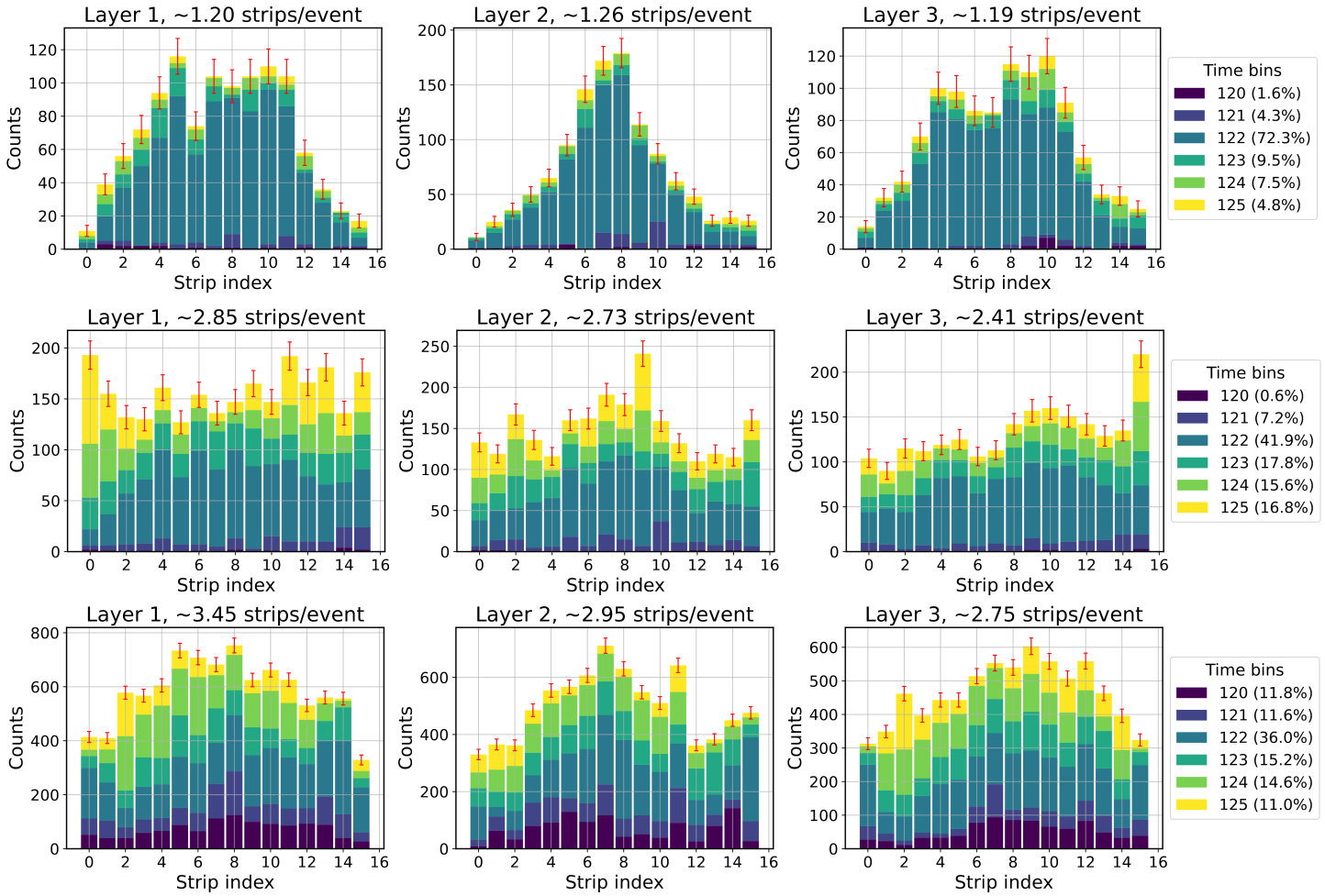}
    \caption{Strip occupancy distributions for the three RPC layers in the D2 cosmic (top), D2 beam (middle), and D3 beam (bottom) datasets. Poisson uncertainties are indicated in red.
}
    \label{D2_D3_occupancies}
\end{figure}
During the D2 and D3 beam tests, all RPC layers were aligned in the same direction. In the D2 cosmic data shown in Figure~\ref{D2_D3_occupancies}, the central strips of the middle layer have higher occupancy than the outer strips, consistent with the angular distribution of cosmic muons. The slight asymmetry in the first and third layers might arise from variations in strip response or nearby shielding.
The D2 cosmic run records approximately one fired strip per layer per event, whereas D2 beam data average 2.9 strips in the first layer and 2.4 in the third. The middle layer shows comparable occupancy in its central and outer strips. Time bin 122 contains 72\% of cosmic entries, compared with 42\% and 39\% in D2 and D3 beam data, respectively, indicating broader timing distributions during beam operation.
Figure~\ref{D2_D3_mutiplicity} shows mean strip multiplicities of 3.6, 8.0, and 9.2 for D2 cosmic, D2 beam, and D3 beam data, respectively. All distributions peak at three fired strips, consistent with a substantial cosmic-like contribution. D2 beam exhibits an extended high-multiplicity tail, while D3 beam shows a discontinuous distribution reaching 48 fired strips. Together, the broader timing distributions and higher multiplicities suggest additional beam-related charged-particle activity within the readout window. Simulations are underway to investigate its origin.

\subsection{Angular Distribution}
Muons produced by the beam should follow their direction and appear as an excess of reconstructed tracks near \(\theta_x=0^\circ\), particularly within \(-5^\circ<\theta_x<5^\circ\). The cosmic run provides a reference angular distribution that includes the effects of detector acceptance. Beam activity can also generate unrelated hits that form accidental track candidates.
\begin{figure}[h!]
    \centering

    \includegraphics[width=0.45\linewidth]{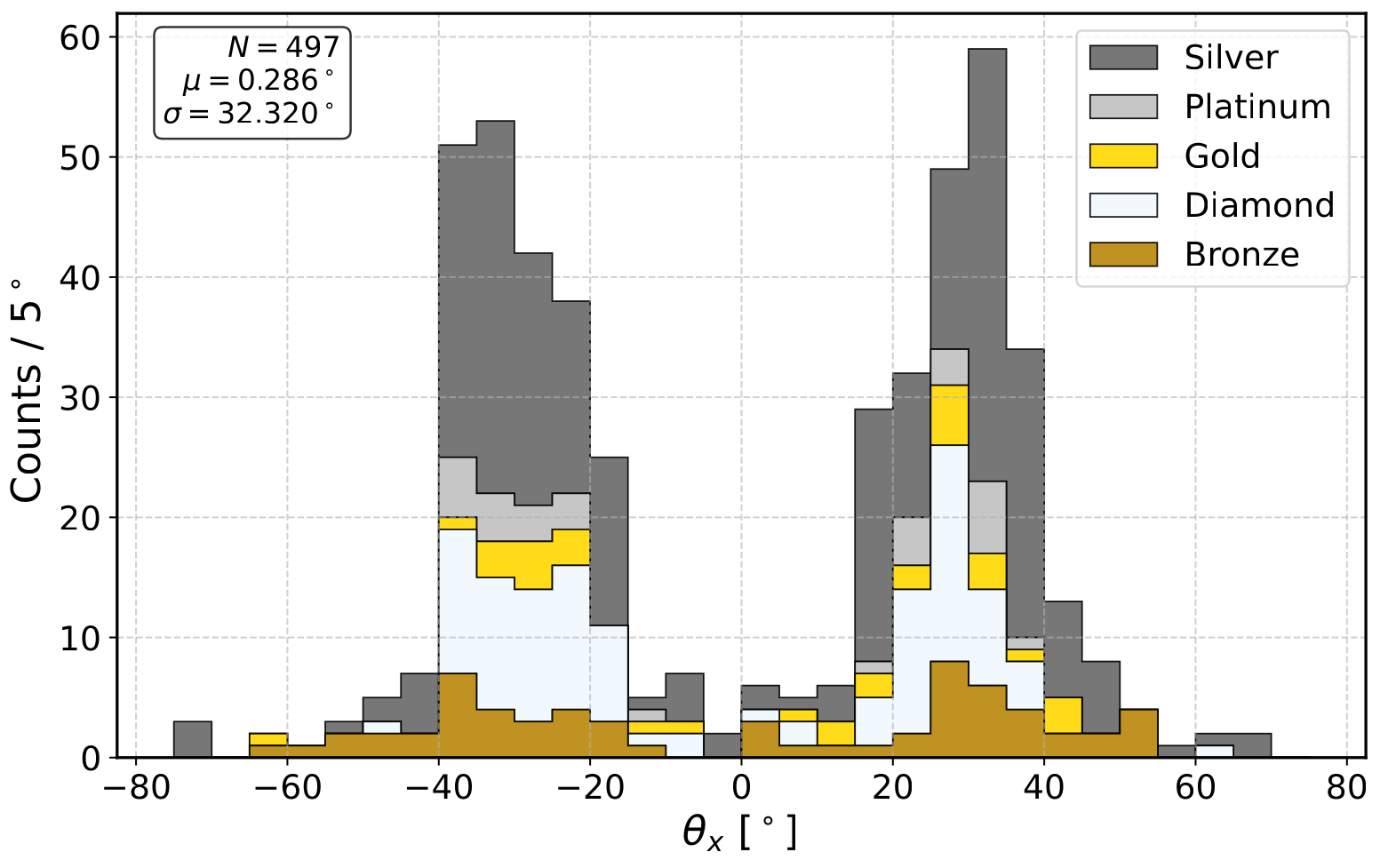}

    \vspace{0.3cm}

    \includegraphics[width=0.45\linewidth]{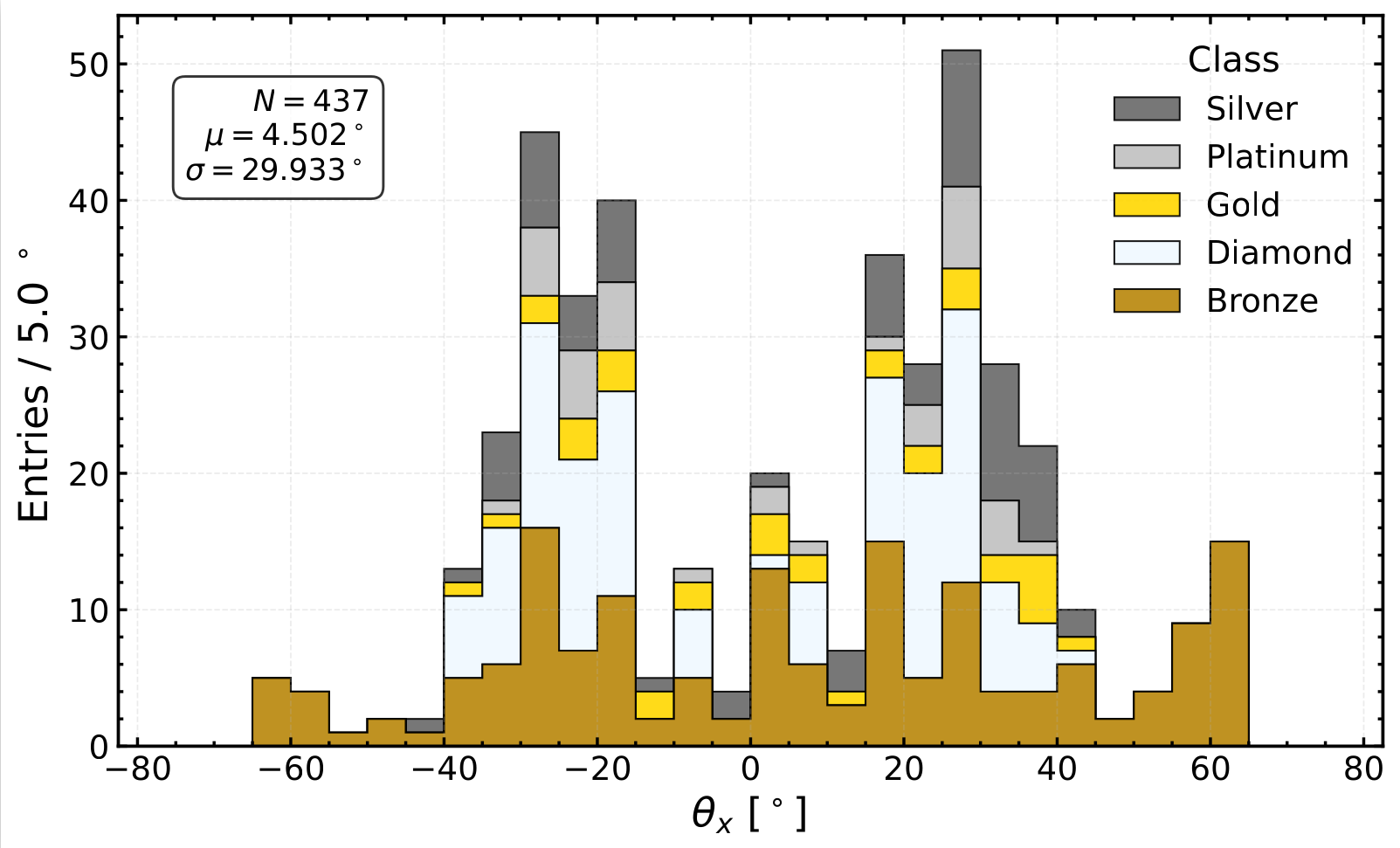}
    \hspace{0.03\linewidth}
    \includegraphics[width=0.45\linewidth]{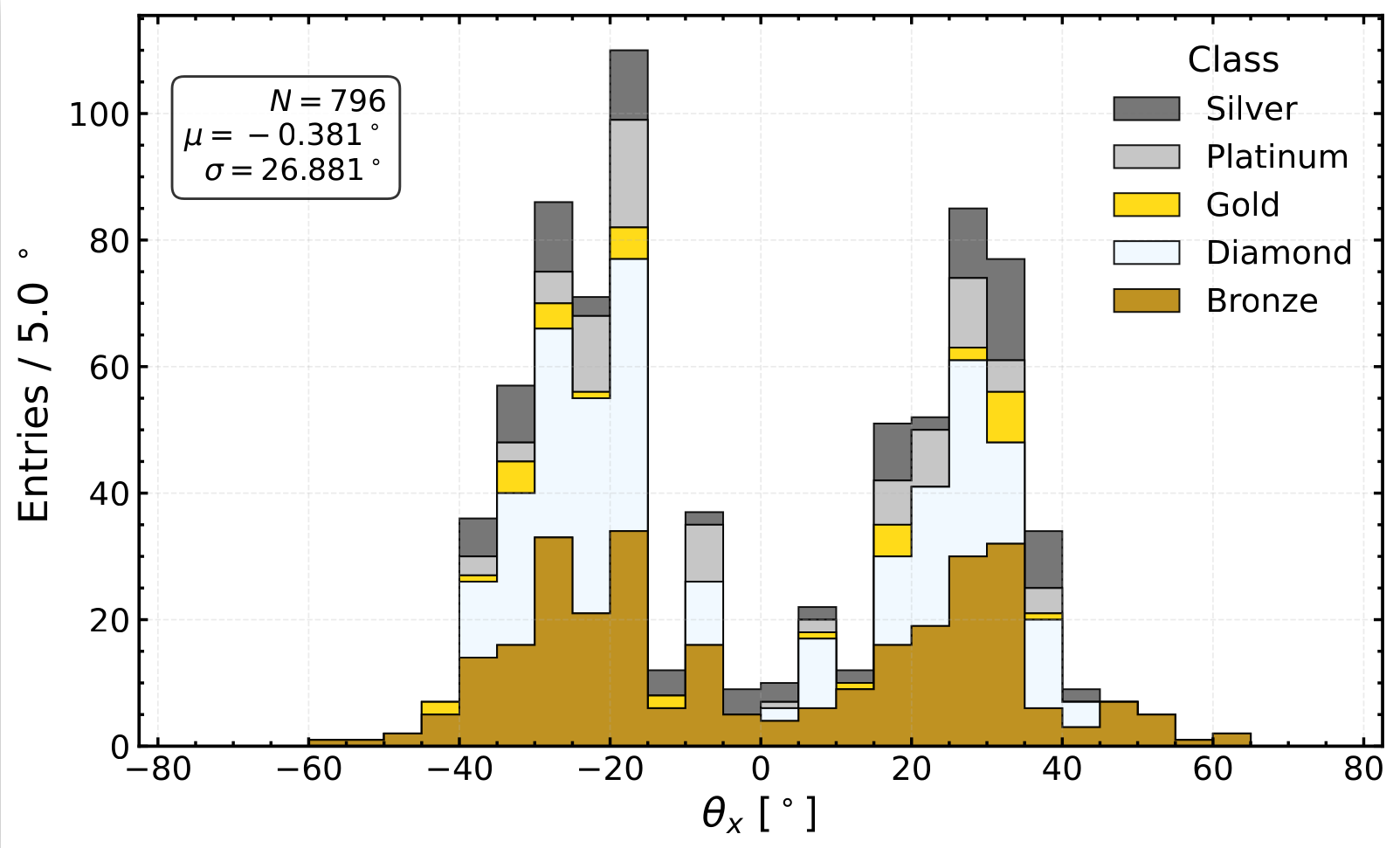}

    \caption{Distributions of the angle $\theta_x$ relative to the
    horizontal plane for the cosmic (top), D2 beam (bottom left),
    and D3 beam (bottom right) datasets. Horizontal and vertical
    muons correspond to $\theta_x = 0^\circ$ and $90^\circ$,
    respectively~\cite{LPA-Ikram2026}.}
    \label{Zenith}
\end{figure}

Figure~\ref{Zenith} shows a broad concentration of tracks within approximately \(-20^\circ<\theta_x<20^\circ\) in the cosmic reference sample. The beam sample follows a similar distribution and shows no distinct excess in the narrower region expected for beam-induced muons. The angular comparison therefore provides no conclusive evidence for a beam-related muon component in the reconstructed tracks.

\subsection{Inter-arrival Time}
\begin{figure}
    \centering
    \includegraphics[width=1\linewidth]{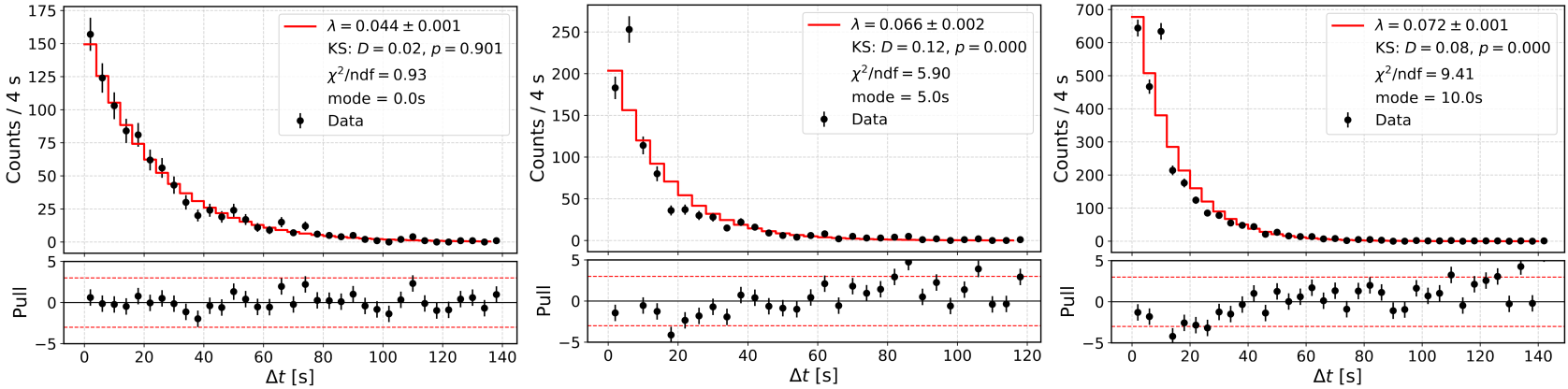}
    \caption{Fits of the time-interval distributions $\Delta t$ for Cosmic (left), D2 beam (center), and D3 beam (right) data using a single-parameter exponential model with rate $\lambda$. The legends report the unbinned $\Delta t$ mode, $\chi^2/\mathrm{ndf}$, and the Kolmogorov–Smirnov statistic $D$ with its corresponding $p$-value. D2 cosmic data are consistent with the exponential model, whereas D2 and D3 beam data show deviations exceeding $5\sigma$ at $\Delta t = 5~s$ and $10~s$, respectively.}
    \label{fig:interarrival_time}
\end{figure}

For an approximately constant cosmic-muon flux and independent arrivals, detections follow a homogeneous Poisson process. If detector dead-time effects are negligible, the interval $\Delta t$ between successive events follows an exponential probability density:
\begin{equation}
    f(\Delta t;\lambda)=\lambda e^{-\lambda\Delta t},
    \qquad \Delta t \geq 0,
\end{equation}
where $\lambda$ is the event rate. For each dataset, we estimate $\hat{\lambda}=1/\langle\Delta t\rangle$ and assess agreement with the exponential model using the Kolmogorov–Smirnov (KS) test and $\chi^2/\mathrm{ndf}$.
 As shown in Figure~\ref{fig:interarrival_time}, the cosmic sample is consistent with the model, yielding $\lambda_{\mathrm{fit,cosmic}}=0.044\pm0.001\:\mathrm~{s}^{-1}$, a KS statistic $D=0.02$, and $p=0.9$.
In contrast, D2 and D3 beam data exhibit pronounced peaks at $\Delta t=5.0~s$ and $10.0~s$, respectively, with local deviations exceeding $5\sigma$. Near-zero KS $p$-values and $\chi^2/\mathrm{ndf}$ values of 5.90 and 9.41 indicate poor agreement with the exponential model. These structures suggest departures from a constant-rate Poisson process, and their temporal modulation corresponds to the repetition rate of the laser during beam operation. The interval distributions alone do not distinguish a varying event rate from correlations between successive arrivals. 

\subsection{PCA Analysis of D3 data}
\label{sec:shielding}

We apply a principal component analysis (PCA)~\cite{Jolliffe2002} to identify patterns in the detector data without assuming a beam or secondary-particle model. PCA is used to support data categorization by identifying dominant patterns and correlations among the measured variables. The analysis incorporates multiplicity, position, and timing information at both the strip and cluster levels. By reducing the dimensionality of these variables, PCA facilitates visualization and interpretation of the data. The input includes every scintillator-triggered event and uses observables calculated before tracking or event selection. Global features describe the number of hit layers and fired strips, along with the intervals to the preceding and following events. Per-layer features record strip and cluster multiplicities, hit presence, and the means and standard deviations of strip position and time bin. Each time bin spans $5~ns$. PCA reduces these correlated measurements to a smaller set of components for comparing event populations.

\begin{figure}
    \centering
    \includegraphics[width=0.5\textwidth]{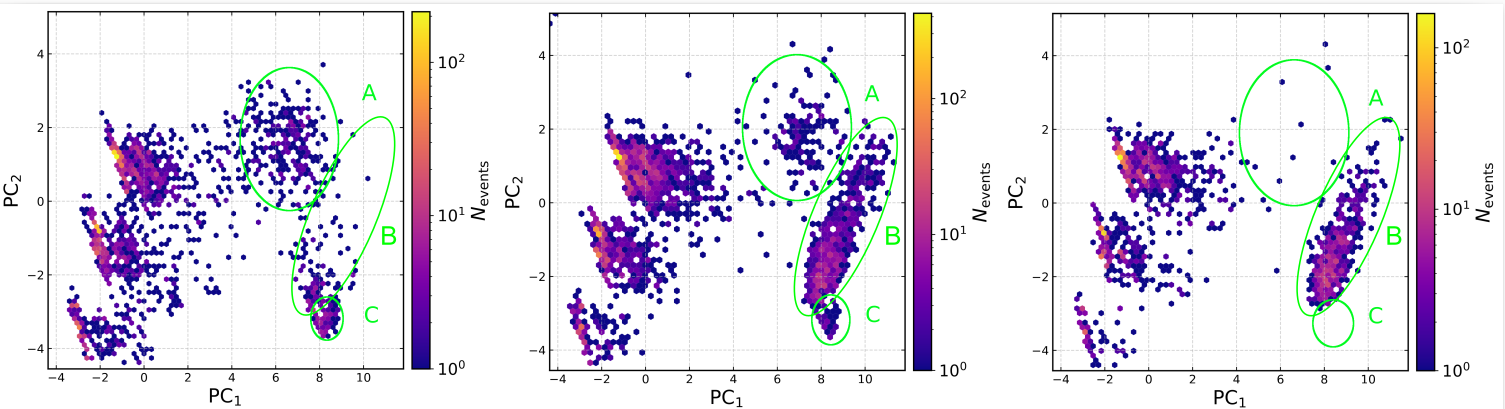}
    \caption{D3 beam data in the PCA plane for three shielding configurations: no shielding (left), lead (center), and lead with polyethylene (right). The distributions are similar across configurations for $\mathrm{PC}_1 < 4$. For $\mathrm{PC}_1 > 4$, distinct groups labeled A, B, and C emerge, each associated with a different range of cluster and strip multiplicities.
}
    \label{D3_PCA}
\end{figure}

To assess how shielding changes the detector response, we compare the event distributions in the $\mathrm{PC}_1$– $\mathrm{PC}_2$ plane. Figure~\ref{D3_PCA} shows distinct populations in the high-activity region, $\mathrm{PC}_1 > 4$. In the lead-only dataset, Region A contains events with more than eight clusters and fewer than 35 fired strips. Region B contains events with more than 35 fired strips and fewer than 12 clusters, while Region C contains events with more than 45 fired strips. Region B also appears with lead and polyethylene (PE) shielding but is largely absent without shielding, suggesting an association with particle production in lead. Region A is more prominent without shielding, whereas Region C appears in both the unshielded and lead-only data but is absent with lead and PE. These differences show that the shielding configuration changes the populations of high-activity events. Identifying the particles responsible for each population requires further study.

\subsection{ Laser Tag}
\begin{figure}[h!]
    \centering

    \begin{tabular}{cc}
        \includegraphics[width=0.45\linewidth]{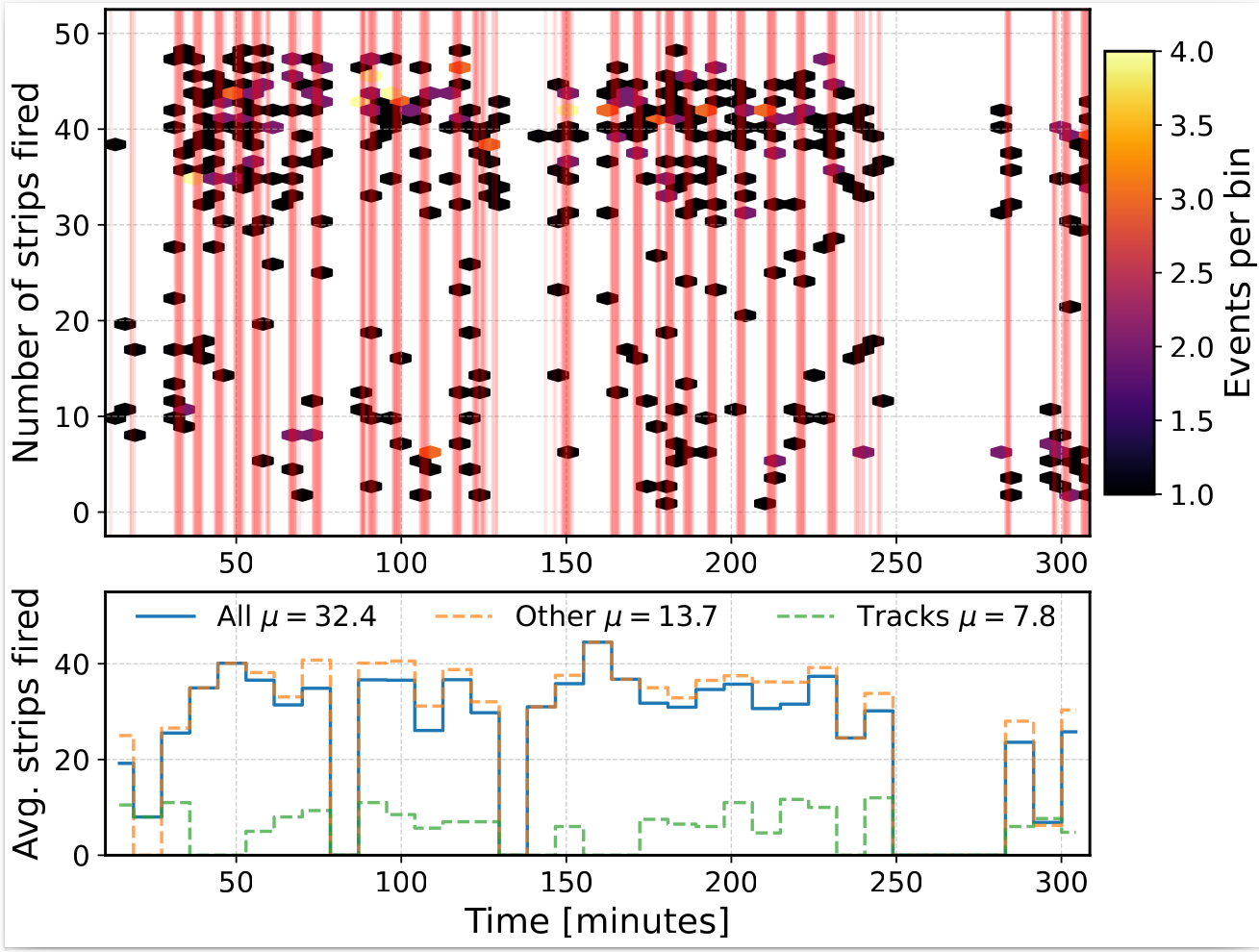} &
        \includegraphics[width=0.45\linewidth]{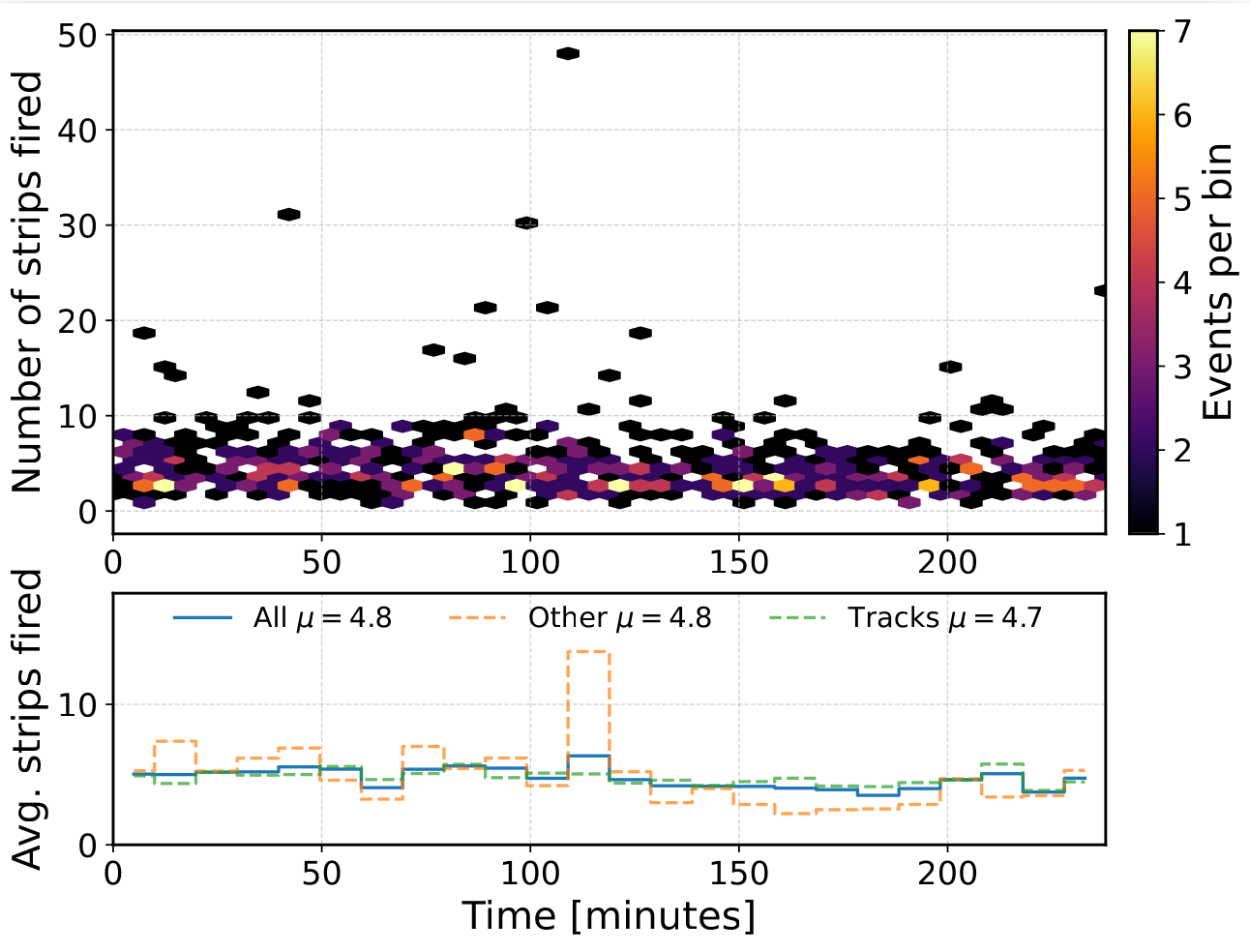} 
          
    \end{tabular}

    \caption{Number of fired strips per event versus the external scintillator trigger time for laser-tagged (left) and non-tagged (right) events. Transparent red lines mark the laser trigger times.}
    \label{laser_tag}
\end{figure}

During the D2 campaign, an additional 5.5-hours run recorded the laser trigger alongside detector data. Events were grouped into laser-tagged and non-tagged samples to compare activity associated with laser operation. Laser timestamps appear as transparent red lines in the strip-multiplicity time series shown in Figure~\ref{laser_tag}.
Detector activity increases during laser-tagged periods. Without a laser tag, tracked events have an average strip multiplicity of 4.8, consistent with the cosmic reference, and high-multiplicity events are rare. During laser-tagged periods, the average strip multiplicity rises to 32.4, indicating substantial additional detector activity associated with beam operation.
These observations support the use of laser tagging to distinguish periods of enhanced beam-related activity from the cosmic-dominated baseline. The correlation supports a beam-related origin for the high-multiplicity events, although it does not establish their particle composition or exclude contributions from detector or acquisition effects.

\section{ Summary and outlook}
\label{sec:summary}

 We operated a portable RPC telescope in sealed mode during parasitic manner at a laser-plasma accelerator. Across these data-taking periods, the telescope provided consistent detector response and tracking information under changing beam conditions. Cosmic-ray reference data and measurements with different shielding configurations helped characterize the backgrounds encountered at Gammatron and ELBA.
The available data do not permit a definitive identification of laser-driven muons. They do, though, establish a practical basis for a dedicated measurement. Future work will combine a laser trigger recorded by the detector DAQ with shielding optimized through particle-transport simulations. These changes will improve background rejection and support a more sensitive search for muons associated with individual laser shots.
\par\vspace{12pt}
\noindent{\bfseries\large Acknowledgments\par}
We thank Ela Rockafellow, Jaron E. Shrock, Howard M. Milchberg, Ari J. Sloss of the University of Maryland laser-plasma group, as well as Sebastian Lorenz and Illia Zymak at ELI Beamlines for providing and operating the electron beam source during the beam test, as documented in Section 2 of Ref.~\cite{LPA-Ikram2026}.
We also thank Mehar Ali Shah, Helio Nogima, Mapse Barroso Ferreira Filho, Salvatore Buontempo, and Kevin Mota Amarilo for their collaboration during the joint data-taking campaigns and for valuable discussions on the interpretation of the data.
We also thank the electronics and mechanical workshops of the CRC at UCLouvain, particularly Quentin de Smedt for helping prepare the detector setup for the beam tests and Nicolas Szilasi for producing the technical drawings of the RPC detectors.
The CP3 team acknowledges financial support from the Fonds de la Recherche Scientifique - FNRS under Grant No. J.0043.22. The VUB team acknowledges the support from the VUB Research Council via Grant No. OZR4149.
\vspace{4pt}
\noindent
\bibliographystyle{JHEP}
\bibliography{references}
\end{document}